\documentclass{revtex4}

\def\Tr{\mathop{\mathbf{Tr}}\nolimits}
\usepackage{graphicx}

\begin{document}

\title{Antinormally Ordered Photodetection of Continuous-mode Field}

\author{Koji Usami$^{\ast,\dagger}$, Akihisa Tomita$^{\ast,\ddagger,\S}$, and Kazuo Nakamura$^{\ast,\dagger,\S}$}
\address{$^{\ast}$Department of Material Science and Engineering, Tokyo Institute of Technology, Yokohama 226-0026, Japan \\
$^{\dagger}$CREST, Japan Science and Technology Agency (JST), Tokyo 150-0002, Japan \\
$^{\ddagger}$ERATO, Japan Science and Technology Agency (JST), Tokyo 113-0033, Japan \\
$^{\S}$Fundamental Research Laboratories, NEC, Tsukuba 305-8501, Japan}

\date{\today}

\begin{abstract}
When the electromagnetic field is detected by stimulated emission, rather than by absorption, antinormally ordered photodetection can be realized. One of the distinct features of this photodetection scheme is its sensitivity to zero-point fluctuation due to the existence of the spontaneous emission. We have recently succeeded in experimentally demonstrating the antinormally ordered photodetection by exploiting nondegenerate stimulated parametric down-conversion process. To properly account for the experiment, the detection process needs to be treated with time-dependent and continuous-mode operators because of the broadband nature of the parametric down-conversion process and the wide spectrum of the pump that we used. Here, we theoretically analyze the antinormally ordered intensity correlation of the continuous-mode fields by pursuing the detection process in the Heisenberg picture. It is shown that the excess positive correlation due to zero-point fluctuation reduces because of the frequency-distinguishability of the two emitted photon pairs.
\end{abstract}

\maketitle


\section{Introduction} \label{sec:intro}

In the context of the quantum theory of optical coherence Glauber found that the measurement operator for a photodetection can be associated with an annihilation operator\cite{Glauber1963}. His assertion rests on the fact that the detection of a photon of energy $\hbar\omega$ is usually carried out with photoelectric effect, which absorbs the photon and dissipates energy $\hbar\omega$ from the field. Any correlations measured by the absorption-based detectors are thus represented by the expectation values of the normally ordered product of the annihilation and the creation operators\cite{Glauber1963}.

The normally ordered photodetections can be said to be insensitive to zero-point fluctuations of the electromagnetic field because the photodetection probability for the vacuum state is zero, i.e., $\Tr[\,|0\rangle \langle 0|\hat{a}^{\dagger}\hat{a}\,]=\langle 0|\hat{a}^{\dagger}\hat{a}|0 \rangle=0$. This insensitivity to zero-point fluctuations results in the photodetection process being \textit{logically irreversible}\cite{UK1992,UIN1996a}, that is, the premeasurement density matrix of the measured system not being uniquely determined by the postmeasurement density matrix and the outcome of the measurement.

When photons are, however, detected by stimulated emission, rather than by absorption, the detection responds not only to actual photons but also to zero-point fluctuations via spontaneous emission. Then, the antinormally ordered photodetection can be realized. Owing to the sensitivity to zero-point fluctuation the antinormally ordered photodetection may provide an interesting alternative approach to continuously monitoring quantum system. This kind of emission-based photodetector, what is called the \textit{quantum counter}, was originally proposed by Bloembergen for detecting infrared photons\cite{Bloembergen1959}. The correlations measured by the quantum counters were then theoretically investigated by Mandel\cite{Mandel1966r_2}. Ueda and Kitagawa showed that the quantum counter, unlike the standard photodetector, make it possible to perform a logically reversible measurement\cite{UK1992}. 

One way to realize the antinormally ordered photodetection is to exploit nondegenerate stimulated parametric down-conversion process. Using this process we have recently succeeded in experimentally implementing the antinormally ordered photodetection and measuring the antinormally ordered intensity correlations for coherent states\cite{UNSTN2004}. The observed correlations were different from the normally ordered ones, as they showed excess positive correlations, i.e., the photon bunching effects, or the Hanbury-Brown$-$Twiss effects\cite{HBT1956n}. 

To properly account for the experiment, the detection process needs to be treated with time-dependent and continuous-mode operators\cite{Loudon} because of the following reasons. First, in the experiment, a ultrashort pulsed field with the broadband spectrum was used as a pump of the parametric down-conversion to obtain large nonlinear response of the crystal and to overcome the slow response time of the photodetectors\cite{KKHM1993,ORW1999}. Second, as a nature of the parametric down-conversion, the spectrum of the down-converted fields themselves were broadband. Moreover, we used a spectral filter for increasing the coherence time of the down-converted field to be larger than that of the pump pulse in order to observe the field correlations. In the previous report\cite{UNSTN2004}, we only presented the simple single-mode treatment of the detection scheme. In this article, we give a theoretical treatment of the antinormally ordered photodetection process by using the continuous-mode field operators and pursuing the detection process in the Heisenberg picture. For the vacuum input, the resultant expression of antinormally ordered correlation is equivalent to that of Ou \textit{et al.}\cite{ORW1999a}, which is obtained by treating the continuous-mode parametric interaction in the Schr\"odinger picture. Our expression for more general coherent field input reasonably explain our experimental result reported previously\cite{UNSTN2004}.

\section{Single-mode treatment} \label{sec:SMT}

\begin{figure}[t]
\begin{center}
\includegraphics[width=0.62\linewidth]{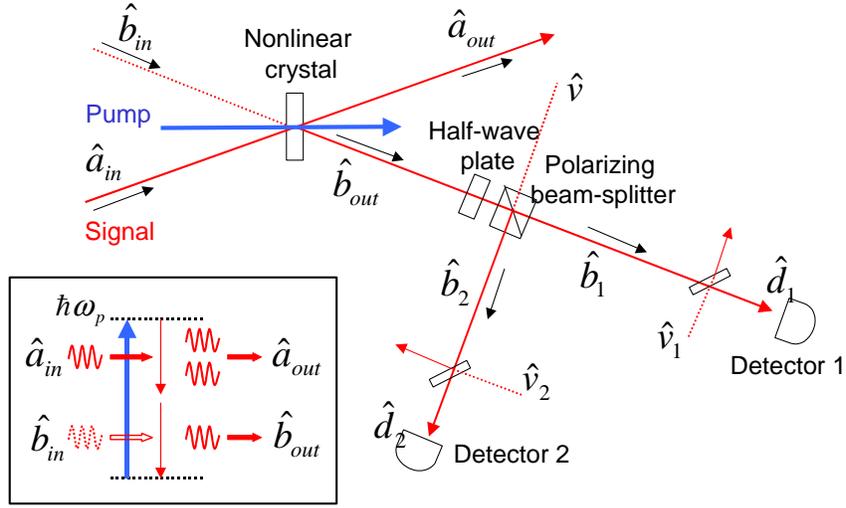}
\caption{Schematic illustration of the antinormally ordered Hanbury-Brown$-$Twiss-type interferometer based on stimulated parametric down-conversion. The inset shows an energy diagram of parametric process.}
\label{fig:schematic}
\end{center}
\end{figure}

The original theoretical proposal of the antinormally ordered photodetection, i.e., the quantum counting\cite{Bloembergen1959} as well as the subsequent theoretical analyses\cite{UK1992,Mandel1966r_2} utilized the cascade emission process of atomic system. This resonant configuration based on the atomic cascade emission can be translated into the non-resonant configuration based on the parametric down-conversion. Our experimental implementation of the antinormally ordered photodetection followed this line. However, to measure the antinormally ordered correlation is by no means a straightforward task because all normally ordered correlation terms should be eradicated. Now, we begin by showing how our devised method\cite{UNSTN2004} works for measuring the antinormally ordered intensity correlation. Here, we will treat the parametric down-conversion process in the simplest way and thus treat only a few single-mode operators relevant in this measurement process. In section~\ref{sec:C_mode}, we will analyze the measurement process with the time-dependent, continuous-mode operators. 
A schematic illustration of the antinormally ordered intensity correlator is shown in Fig.~\ref{fig:schematic}. Now, let us track the time evolution of an annihilation operator, $\hat{a_{in}}$, which represents the signal field. First, operator $\hat{a_{in}}$ is coupled with an operator, $\hat{b_{in}}$, via the parametric interaction with a monochromatic pump field. The coupled spatial differential equations for a mode of the signal field, $\hat{a}(z,\omega)$, and a mode of the auxiliary field, $\hat{b}(z,\omega_{p}-\omega)$, in the nonlinear dielectric medium with the monochromatic pump field of frequency $\omega_{p}$ are written as
\begin{eqnarray}
&\ &\frac{\partial \,\hat{b}(z,\omega_{p}-\omega)}{\partial z} = -s(\omega,\omega_{p})\,e^{i\vartheta(\omega,\omega_{p})} \,\hat{a}^{\dagger}(z,\omega) \ e^{i\Delta k(\omega_{p},\omega)z}, \label{eq:PI1} \\
&\ &\frac{\partial \,\hat{a}^{\dagger}(z,\omega)}{\partial z} = -s(\omega_{p}-\omega,\omega_{p}) \,e^{-i\vartheta(\omega_{p}-\omega,\omega_{p})} \,\hat{b}(z,\omega_{p}-\omega) \ e^{-i\Delta k(\omega_{p},\omega)z}. \label{eq:PI2} 
\end{eqnarray}
where the phase-matching parameter, $\Delta k(\omega_{p},\omega)$, is defined by
\begin{equation}
\Delta k(\omega_{p},\omega)\equiv \frac{\omega_{p}n_{\omega_{p}}}{c}-\frac{\omega n_{\omega}}{c}-\frac{[\omega_{p}-\omega]n_{\omega_{p}-\omega}}{c}.  \label{eq:phasematch}
\end{equation}
Here, $n_{\omega_{p}}$, $n_{\omega}$, and $n_{\omega_{p}-\omega}$ are the refractive indices of the nonlinear dielectric medium at pump field frequency $\omega_{p}$, signal field frequency $\omega$, and auxiliary field frequency $\omega_{p}-\omega$, respectively. $s(\omega,\omega_{p})$ and $\vartheta(\omega,\omega_{p})$ are explicitly given by\cite{Loudon}  
\begin{equation}
s(\omega,\omega_{p})\,e^{i\vartheta(\omega,\omega_{p})}=-\sqrt{\frac{F(\omega_{p})\hbar \omega_{p} \omega [\omega_{p}-\omega]}{8\epsilon_{0}c^{3}A\,n_{\omega_{p}}n_{\omega}n_{\omega_{p}-\omega}}}\chi^{(2)}(\omega,\omega_{p})e^{i\phi(\omega_{p})}, \label{eq:gain}
\end{equation}
and considered as the coupling constant between the two modes, $\hat{a}(z,\omega)$ and $\hat{b}(z,\omega_{p}-\omega)$. Here, $F(\omega_{p})$ and $\phi(\omega_{p})$ correspond to the power spectral density and the phase of the pump field, respectively. $\chi^{(2)}(\omega,\omega_{p})$ is the second-order nonlinear susceptibility. We can easily show that $s(\omega,\omega_{p})\,e^{i\vartheta(\omega,\omega_{p})} = s(\omega_{p}-\omega,\omega_{p})\,e^{i\vartheta(\omega_{p}-\omega,\omega_{p})}$. The two modes, $\hat{a}(z,\omega)$ and $\hat{b}(z,\omega_{p}-\omega)$, as well as the pump field are assumed to be propagating almost colinearly in the $z$ direction, here. When the phase-matching condition is perfectly satisfied, i.e., $\Delta k(\omega_{p},\omega)=0$, we have the following solutions:
\begin{eqnarray}
\hat{a_{out}} &\equiv& \hat{a}(L,\omega) = \hat{a_{in}} \cosh [s L]-\hat{b_{in}}^{\dagger} e^{i \vartheta} \sinh [s L] \nonumber \\
&=& \hat{S}_{ab}^{\dagger}(\zeta)\hat{a_{in}}\hat{S}_{ab}(\zeta) \label{eq:parametric1} \\
\hat{b_{out}} &\equiv& \hat{b}(L,\omega_{p}-\omega) = \hat{b_{in}} \cosh [s L]-\hat{a_{in}}^{\dagger} e^{i \vartheta} \sinh [s L] \nonumber \\
&=& \hat{S}_{ab}^{\dagger}(\zeta)\hat{b_{in}}\hat{S}_{ab}(\zeta) \label{eq:parametric2}
\end{eqnarray}
with the initial conditions $\hat{a}(0,\omega)\equiv\hat{a_{in}}$ and $\hat{b}(0,\omega_{p}-\omega)\equiv\hat{b_{in}}$. Here, $L$ is the length of the crystal. $\hat{S}_{ab}$ is the two-mode squeezing operator, which is given by
\begin{equation} \label{eq:TMsqueezed}
\hat{S}_{ab}(\zeta) = \exp[\,\zeta^{*} \hat{a_{in}}\hat{b_{in}}-\zeta \hat{a_{in}}^{\dagger}\hat{b_{in}}^{\dagger}]
\end{equation}
with $\zeta \equiv s L \exp[i\vartheta]$. Intuitively, one photon in the pump field is split into two photons in such a way that the total energy of the fields is conserved during the process. An energy diagram of this process is shown in the inset of Fig.~\ref{fig:schematic}. 

Next, by dividing the field represented by $\hat{b_{out}}$ into fields $\hat{b_{1}}$ and $\hat{b_{2}}$ with a half-wave plate and a polarizing beam-splitter, the standard Hanbury-Brown$-$Twiss interferometer\cite{HBT1956n} for field $\hat{b_{out}}$ is formed as shown in Fig.~\ref{fig:schematic}. Here, the imperfect detectors and several optical losses are taken into account by introducing auxiliary beam-splitters with vacuum fields $\hat{v_{1}}$ and $\hat{v_{2}}$\cite{Loudon}. Output fields $\hat{d_{1}}$ and $\hat{d_{2}}$ of Fig.~\ref{fig:schematic} are then written as
\begin{eqnarray}
\hat{d_{1}} = \sqrt{\eta_{1}}\ [\mathcal{T} \,\hat{b_{out}}+\mathcal{R} \,\hat{v}] + i \sqrt{1-\eta_{1}} \,\hat{v_{1}}, \nonumber \\
\hat{d_{2}} = \sqrt{\eta_{2}}\ [\mathcal{R} \,\hat{b_{out}}+\mathcal{T} \,\hat{v}] + i \sqrt{1-\eta_{2}} \,\hat{v_{2}},
\label{eq:output}
\end{eqnarray}
respectively. Here, $\eta_{1}$ and $\eta_{2}$ are the total photodetection efficiencies for detectors~1 and 2, respectively; $\mathcal{R}$ and $\mathcal{T}$ denote the reflection and transmission coefficients at the beam-splitter, respectively, and can be varied with the half-wave plate and the polarizing beam-splitter; and $\hat{v}$ is an auxiliary vacuum field introduced from the empty port of the beam-splitter.

Since the modes relevant to operators $\hat{b_{in}}$, $\hat{v}$, $\hat{v_{1}}$, and $\hat{v_{2}}$ are initially vacua, the surviving contribution to the number of photodetection events of the normally ordered photodetection for fields $\hat{d_{1}}$ and $\hat{d_{2}}$ coincides with that of the antinormally ordered photodetection for field $\hat{a_{in}}$ up to a constant factor, i.e., 
\begin{eqnarray}
\langle \hat{n}_{d_{1}} \rangle &\equiv& \langle \hat{d_{1}}^{\dagger} \hat{d_{1}} \rangle = \eta_{1}|\mathcal{T}|^{2}\sinh^{2} [s L] \langle \hat{a_{in}} \hat{a_{in}}^{\dagger} \rangle, \nonumber \\
\langle \hat{n}_{d_{2}} \rangle &\equiv& \langle \hat{d_{2}}^{\dagger} \hat{d_{2}} \rangle = \eta_{2}|\mathcal{R}|^{2}\sinh^{2} [s L] \langle \hat{a_{in}} \hat{a_{in}}^{\dagger} \rangle,
\label{eq:single}
\end{eqnarray}
respectively. Furthermore, the number of coincidental photodetection events of fields $\hat{d_{1}}$ and $\hat{d_{2}}$ (i.e., the interbeam correlation for the two fields) results in 
\begin{equation}
\langle \hat{n}_{d_{1}} \hat{n}_{d_{2}} \rangle \equiv \langle \hat{d_{1}}^{\dagger} \hat{d_{1}} \hat{d_{2}}^{\dagger} \hat{d_{2}} \rangle = \eta_{1}\eta_{2}|\mathcal{T}|^{2}|\mathcal{R}|^{2}\sinh^{4} [s L]  \langle \hat{a_{in}} \hat{a_{in}} \hat{a_{in}}^{\dagger} \hat{a_{in}}^{\dagger} \rangle,
\label{eq:Acorrelation}
\end{equation}
where we use the commutation relation for each operator and relations $\mathcal{TR^{*}}=-\mathcal{RT^{*}}$, which should be satisfied when the lossless beam splitter is used. Thus, the surviving contribution to the coincidences turns out to be made only by operators $\hat{a_{in}}$ and $\hat{a_{in}}^{\dagger}$ in antinormal order. Consequently, we can evaluate the antinormally ordered intensity correlation for field $\hat{a_{in}}$ as follows:
\begin{equation}
g^{(2)}_{1,2} \equiv \frac{\langle \hat{n}_{d_{1}} \hat{n}_{d_{2}} \rangle}{\langle \hat{n}_{d_{1}} \rangle \langle \hat{n}_{d_{2}} \rangle} = \frac{\langle \hat{a_{in}} \hat{a_{in}} \hat{a_{in}}^{\dagger} \hat{a_{in}}^{\dagger} \rangle}{\langle \hat{a_{in}} \hat{a_{in}}^{\dagger} \rangle \langle \hat{a_{in}} \hat{a_{in}}^{\dagger} \rangle} = g^{(2[A])}.  \label{eq:ANHBT_e}
\end{equation}
Note that Eq.~(\ref{eq:ANHBT_e}) holds regardless of the splitting ratio at the beam-splitter, the quantum efficiencies of the detectors, and the optical losses. In this way, we can realize the measurement of the antinormally ordered intensity correlation by exploiting the parametric interaction, Eqs.~(\ref{eq:parametric1}) and (\ref{eq:parametric2}).

\section{Continuous-mode treatment} \label{sec:C_mode}

To analyze the correlation observed in our experiment, the detection process should be treated with time-dependent and continuous-mode operators\cite{Loudon}. In this treatment, the measured antinormally ordered intensity correlations in Eq.~(\ref{eq:ANHBT_e}) becomes time dependent. Moreover, since the response-time jitter of the detector is larger than the pump pulse duration (the duration of the parametric interaction) but smaller than the time interval between two successive pulses, the relevant information on the time dependence is embodied in the integrated number of the delayed coincidences over the response-time jitter of the detectors\cite{KKHM1993,ORW1999}. In the following we treat the antinormally ordered photodetection process by using the continuous-mode field operators and pursuing the detection process in the Heisenberg picture. For the vacuum input, the resultant expression of antinormally ordered correlation is equivalent to that of Ou \textit{et al.}\cite{ORW1999a}, which is derived by the continuous-mode analysis of the parametric interaction in the Schr\"odinger picture. 

We start by reconsidering the quantization procedure to deal with the realistic photodetection process, in which a photon in a \textit{travelling field} with a broadband spectrum is dissipated at a detector's photosensitive surface (cross-section $A$), during period $T$. Then, we tackle the continuous-mode parametric interaction in the Heisenberg picture. Finally, the antinormally ordered intensity correlations of continuous-mode fields are derived.

\subsection{Quantization procedure for travelling field} \label{sec:QP4TF}

Following the method developed by Huttner, Serulnik, and Ben-Aryeh\cite{HSB1990a}, we review a quantization procedure for treating the travelling field with multimode operators. An annihilation (a creation) operator, which annihilates (creates) a photon of frequency $\omega_{m}$ travelling through plane $z=z_{0}$ with cross-section $A$ during period $T$, is defined as $\hat{a}(z_{0},\omega_{m})$ ($\hat{a}^{\dagger}(z_{0},\omega_{m})$). Here $\omega_{m}=2\pi m/T$. The positive-frequency parts of electric and magnetic fields in vacuum are written as $\hat{E}^{(+)}(z_{0},t)=\sum_{m} \sqrt{\frac{\hbar \omega_{m}}{2\epsilon_{0}AcT}} \hat{a}(z_{0},\omega_{m})e^{-i\omega_{m}t}$ and $\hat{B}^{(+)}(z_{0},t)=\sum_{m} \sqrt{\frac{\hbar \omega_{m}}{2\epsilon_{0}Ac^{3}T}} \hat{a}(z_{0},\omega_{m})e^{-i\omega_{m}t}$, respectively, where $\hbar$ is the Planck's constant, $\epsilon_{0}$ is the electric permittivity of vacuum, and $c$ is the speed of light in vacuum. Then the normally ordered Poynting vector\cite{Loudon,HSB1990a} takes the form
\begin{eqnarray}
\hat{S}(z_{0},t) &=& c^{2}\epsilon_{0} [\hat{E}^{(-)}(z_{0},t)\hat{B}^{(+)}(z_{0},t)+\hat{B}^{(-)}(z_{0},t)\hat{E}^{(+)}(z_{0},t)] \nonumber \\
&=& \sum_{m,m'}\frac{\hbar}{2AT}\sqrt{\omega_{m}\omega_{m'}}[\hat{a}^{\dagger}(z_{0},\omega_{m})\hat{a}(z_{0},\omega_{m'})e^{i[\omega_{m}-\omega_{m'}]t}+H.c.]. \label{eq:Poynting}
\end{eqnarray}
Here and throughout this subsection, we exclusively use $()$ for denoting variables to avoid any confusion. By integrating the Poynting vector over the period, $T$, we have the total energy, 
\begin{equation}
A\int^{T}_{0}dt\hat{S}(z_{0},t)
= \sum_{m}\hbar\omega_{m} \hat{a}^{\dagger}(z_{0},\omega_{m})\hat{a}(z_{0},\omega_{m}), \label{eq:energy}
\end{equation}
which flows through the plane, $z=z_{0}$, with a cross-section, $A$. The quantization volume for the operators $\hat{a}(z_{0},\omega_{m})$ and $\hat{a}^{\dagger}(z_{0},\omega_{m})$ is thus $AcT$. It is important to note that the total energy, Eq.~(\ref{eq:energy}), which is calculated without assuming a real cavity, does not involve zero-point energy. Thus the total energy, Eq.~(\ref{eq:energy}), can be considered as the energy that is dissipated at a detector's photosensitive surface at $z=z_{0}$ with cross-section $A$, during the period, $T$. Here, the equal-space commutation relations (ESCR):
\begin{equation}
[\hat{a}(z_{0},\omega_{m}), \hat{a}^{\dagger}(z_{0},\omega_{m'})]=\delta_{m,m'} \label{eq:ESCR}
\end{equation}
are introduced\cite{HSB1990a}. It can be verified that these ESCR are consistent with the usual equal-time commutation relations (ETCR) of the canonical quantization procedure and are even valid for the case of a field in a dielectric medium\cite{HSB1990a,HB1992a}.

The number operator of photons localized in the finite volume, $AcT$, can be then written as
\begin{equation}
\hat{N}(z_{0})=\sum_{m} \hat{a}^{\dagger}(z_{0},\omega_{m})\hat{a}(z_{0},\omega_{m})=\int^{T}_{0}dt\hat{I}(z_{0},t), \label{eq:number}
\end{equation}
where the operator, $\hat{I}(z_{0},t)$, represents the photon flux at position $z=z_{0}$ and cross-section $A$. The photon flux operator, $\hat{I}(z_{0},t)$, takes the simple form:
\begin{equation}
\hat{I}(z_{0},t)=\hat{d}^{\dagger}(z_{0},t)\hat{d}(z_{0},t).  \label{eq:flux}
\end{equation}
with the photon flux amplitude operator,
\begin{equation}
\hat{d}(z_{0},t)=\sum_{m} \frac{1}{\sqrt{T}} \hat{a}(z_{0},\omega_{m})e^{-i\omega_{m}t}.  \label{eq:fluxA}
\end{equation}
Here, we stress that even though the number operator, Eq.~(\ref{eq:number}), is taken at position $z=z_{0}$, it is not the number of strictly localized photons at $z=z_{0}$, rather it means the number of delocalized photons over the length, $cT$, and the cross-section, $A$; otherwise a localized number operator for photons would contradict the well-known fact that there does not exist a probability density for the position of photons\cite{Cook1982a,CDG,Mandel1966r_1}. 

The continuous-mode extension of the above multimode treatment is straightforward by taking the limit, $T \to \infty$, with almost-always-valid narrow-bandwidth approximation, by which the range of integration over frequency $\omega$ is extended to cover from $-\infty$ to $\infty$\cite{Loudon}. For instance, the photon flux amplitude operator, Eq.~(\ref{eq:fluxA}), becomes $\hat{d}(z_{0},t)=\frac{1}{\sqrt{2\pi}}\int_{-\infty}^{\infty} d\omega \hat{a}(z_{0},\omega)e^{-i\omega t}$.

Hereafter, we will work with the continuous-mode extension of the annihilation and the creation operators, which satisfy the continuous-mode ESCR, 
\begin{equation}
[\hat{a}(z_{0},\omega), \hat{a}^{\dagger}(z_{0},\omega')]=\delta(\omega,\omega'), \label{eq:ESCR2}
\end{equation}
which is analogous to the multimode ESCR, Eq.~(\ref{eq:ESCR}). To include the effects of the spectral filter with the amplitude transmission function, $\xi(\omega)$, we can further modify the photon flux amplitude operator, Eq.~(\ref{eq:fluxA}), as
\begin{equation}
\hat{d_{\xi}}(z_{0},t)=\frac{1}{\sqrt{2\pi}}\int_{-\infty}^{\infty} d\omega \xi(\omega) \hat{a}(z_{0},\omega)e^{-i\omega t}.  \label{eq:fluxA3}
\end{equation}
Then, the number operator, Eq.~(\ref{eq:number}), is changed to
\begin{eqnarray} 
\hat{N}(z_{0}) &=& \int_{\infty}^{\infty}dt\hat{I}(z_{0},t) = \int_{-\infty}^{\infty}dt \hat{d}^{\dagger}(z_{0},t)\hat{d}(z_{0},t) \nonumber \\
&=& \int_{-\infty}^{\infty}d\omega|\xi(\omega)|^{2} \  \hat{a}^{\dagger}(z_{0},\omega)\hat{a}(z_{0},\omega). \label{eq:number2}
\end{eqnarray}
The operator, $\hat{N}(z_{0})$, can be identified as a \textit{positive operator-valued measure}\cite{NC} for a photodetection measurement with the spectral filter characterized by $\xi(\omega)$ during infinite time. This expression is an approximation of the rather slow response time of the standard photodetector and indicates that the intermode coherences disappear during the photodetection\cite{ALS2002epjd}. 

\subsection{Continuous-mode analysis of parametric down-conversion in the Heisenberg picture} \label{sec:C_modePDC}

In Sec.~\ref{sec:SMT} the analysis of parametric interaction was carried out under the perfect phase-matching condition, i.e., $\Delta k(\omega_{p},\omega)=0$. Besides, the pump field was taken as to be monochromatic. In our experiment\cite{UNSTN2004}, however, the condition, $\Delta k(\omega_{p},\omega)=0$, did not perfectly satisfy in generating the down-converted fields, and the pump field was a ultrashort pulsed field with the broadband spectrum. In this subsection, to appropriately examine the experiment we will develop a theory of the stimulated parametric down-conversion process for the continuous-mode fields in the Heisenberg picture. 

To elaborate the theory of the parametric process so as to include the pump field spectrum and the frequency spread of the down-converted fields, let us go back to the two-mode squeezing operator, $\hat{S}_{ab}$ of Eq.~(\ref{eq:TMsqueezed}). The continuous-mode expansion of the squeezing operator can be given by
\begin{equation} \label{eq:CMsqueezed}
\hat{S}_{c} = \exp[\,\hat{P}_{f} -\hat{P}_{f}^{\dagger}],
\end{equation}
where $\hat{P}_{f}^{\dagger}$ may be called the \textit{frequency-entangled photon-pair creation operator}, which is defined by
\begin{equation} \label{eq:CMppC}
\hat{P}_{f}^{\dagger}= \int_{-\infty}^{\infty} d\omega_{1} d\omega_{2} \tilde{\zeta}(\omega_{1},\omega_{2})\phi(\omega_{1},\omega_{2}) \hat{a_{in}}^{\dagger}(\omega_{1})\hat{b_{in}}^{\dagger}(\omega_{2}),
\end{equation}
where
\begin{equation} \label{eq:zeta}
\tilde{\zeta}(\omega_{1},\omega_{2}) \equiv s(\omega_{1},\omega_{1}+\omega_{2}) L \exp[i\vartheta(\omega_{1},\omega_{1}+\omega_{2})].
\end{equation}
Since the refractive indices, $n_{\omega_{1}+\omega_{2}}$, $n_{\omega_{1}}$, and $n_{\omega_{2}}$, the frequency multiple, $(\omega_{1}+\omega_{2})\omega_{1}\omega_{2}$, and the second-order nonlinear susceptibility, $\chi^{(2)}(\omega,\omega_{p})$, in $s(\omega_{1},\omega_{1}+\omega_{2})$, which was given by Eq.~(\ref{eq:gain}), can be approximated by constants with respect to frequencies $\omega_{1}$ and $\omega_{1}$ over the spectral filter's bandwidth (which is determined by the profile, $|\xi(\omega)|^{2}$ in Eq.~(\ref{eq:number2})), they can thus be factored out as
\begin{equation} \label{eq:zeta2}
\tilde{\zeta}(\omega_{1},\omega_{2}) \approx \zeta \sqrt{F_{p}(\omega_{1}+\omega_{2})}.
\end{equation}
The remaining frequency-dependent terms of $\hat{P}_{f}^{\dagger}$ in Eq.~(\ref{eq:CMppC}) are the pump field profile, $F_{p}(\omega_{1}+\omega_{2})$, and the phase-mismatching function, $\phi(\omega_{1},\omega_{2})$. Then, the continuous-mode counterparts of the down-converted fields, Eqs.~(\ref{eq:parametric1}) and (\ref{eq:parametric2}), can be written as
\begin{eqnarray}
\hat{a_{out}}(\omega) &=& \hat{S}_{c}^{\dagger}(\zeta)\hat{a_{in}}(\omega)\hat{S}_{c}(\zeta), \label{eq:CMparametric1} \\
\hat{b_{out}}(\omega) &=& \hat{S}_{c}^{\dagger}(\zeta)\hat{b_{in}}(\omega)\hat{S}_{c}(\zeta), \label{eq:CMparametric2}
\end{eqnarray}
respectively. Here, $\hat{S}_{c}(\zeta)$ is defined by
\begin{equation} \label{eq:CMsqueezed_2}
\hat{S}_{c}(\zeta) = \exp[\,\zeta^{*}\hat{P}_{\Phi} -\zeta \hat{P}_{\Phi}^{\dagger}],
\end{equation}
with
\begin{eqnarray} 
\hat{P}_{\Phi}^{\dagger} &=& \int_{-\infty}^{\infty} d\omega_{1} d\omega_{2} \sqrt{F_{p}(\omega_{1}+\omega_{2})}\,\phi(\omega_{1},\omega_{2}) \hat{a_{in}}^{\dagger}(\omega_{1})\hat{b_{in}}^{\dagger}(\omega_{2}) \nonumber \\
&\equiv& \int_{-\infty}^{\infty} d\omega_{1} d\omega_{2} \Phi(\omega_{1},\omega_{2}) \hat{a_{in}}^{\dagger}(\omega_{1})\hat{b_{in}}^{\dagger}(\omega_{2}). \label{eq:CMppC_2}
\end{eqnarray}

\subsection{Antinormally ordered intensity correlation} \label{sub:aHBT_STDC_C}

By incorporating the results of two subsections above, we will now show how the antinormally ordered intensity correlation of single-mode field given by Eq.~(\ref{eq:ANHBT_e}) is altered when the field is replaced with continuous-mode field. First of all, the output operators, $\hat{d_{1}}$ and $\hat{d_{2}}$, of Eq.~(\ref{eq:output}) are rewritten as the flux amplitudes,
\begin{eqnarray}
\hat{d_{1}}(t) &=& \frac{1}{\sqrt{2\pi}}\int_{-\infty}^{\infty} d\omega \xi(\omega) \hat{d_{1}}(\omega)e^{-i\omega t}, \nonumber \\
\hat{d_{2}}(t) &=& \frac{1}{\sqrt{2\pi}}\int_{-\infty}^{\infty} d\omega \xi(\omega) \hat{d_{2}}(\omega)e^{-i\omega t}, \label{eq:output1}
\end{eqnarray}
with
\begin{eqnarray}
\hat{d_{1}}(\omega) &=& \sqrt{\eta_{1}}\ [\mathcal{T} \,\hat{b_{out}}(\omega)+\mathcal{R} \,\hat{v}(\omega)] + i \sqrt{1-\eta_{1}} \,\hat{v_{1}}(\omega), \nonumber \\
\hat{d_{2}}(\omega) &=& \sqrt{\eta_{2}}\ [\mathcal{R} \,\hat{b_{out}}(\omega)+\mathcal{T} \,\hat{v}(\omega)] + i \sqrt{1-\eta_{2}} \,\hat{v_{2}}(\omega),
\label{eq:output2}
\end{eqnarray}
respectively, where $\hat{b_{out}}(\omega)$ was given by Eq.~(\ref{eq:CMparametric2}). Here, the parameters, $\mathcal{T}$, $\mathcal{R}$, $\eta_{1}$, and $\eta_{2}$, are assumed to be independent of the frequencies of the fields. As noted in Sec.~\ref{sec:SMT}, the modes relevant to operators $\hat{v}(\omega)$, $\hat{v_{1}}(\omega)$, and $\hat{v_{2}}(\omega)$ are supposed to be initially vacua. Then, the number of the photodetection events for the field $\hat{d_{1}}(\omega)$ with detector~1 reduces to
\begin{eqnarray}
&\!&\langle \hat{N}_{d_{1}} \rangle = \int_{-\infty}^{\infty} dt \big\langle \hat{d_{1}}^{\dagger}(t) \hat{d_{1}}(t) \big\rangle = \frac{1}{2 \pi}\int_{-\infty}^{\infty} dt \int_{-\infty}^{\infty} \!\!\!d\omega d\omega' \xi^{*}(\omega)\xi(\omega') \big\langle \hat{d_{1}}^{\dagger}(\omega) \hat{d_{1}}(\omega') \big\rangle \nonumber \\
&=& \frac{1}{2\pi}\eta_{1}|\mathcal{T}|^{2} \!\int_{-\infty}^{\infty} \!\!\!dt \int_{-\infty}^{\infty} \!\!\!d\omega d\omega' \xi^{*}(\omega)\xi(\omega') \big\langle \hat{b_{out}}^{\dagger}(\omega) \hat{b_{out}}(\omega') \big\rangle e^{i(\omega-\omega')t} \nonumber \\
&=& \frac{1}{2\pi}\eta_{1}|\mathcal{T}|^{2} \!\int_{-\infty}^{\infty} \!\!\!dt \int_{-\infty}^{\infty} \!\!\!d\omega d\omega' \xi^{*}(\omega)\xi(\omega') \big\langle \hat{S}_{c}^{\dagger}(\zeta)\hat{b_{in}}^{\dagger}(\omega) \hat{S}_{c}(\zeta) \hat{S}_{c}^{\dagger}(\zeta) \hat{b_{in}}(\omega')\hat{S}_{c}(\zeta) \big\rangle e^{i(\omega-\omega')t} \nonumber \\
&=& \frac{1}{2\pi}\eta_{1}|\mathcal{T}|^{2} \!\int_{-\infty}^{\infty} \!\!\!dt \int_{-\infty}^{\infty} \!\!\!d\omega d\omega' \xi^{*}(\omega)\xi(\omega') \big\langle \hat{S}_{c}^{\dagger}(\zeta)\hat{b_{in}}^{\dagger}(\omega) \hat{b_{in}}(\omega')\hat{S}_{c}(\zeta) \big\rangle e^{i(\omega-\omega')t}, \label{eq:singleM1}
\end{eqnarray}
where we use Eq.~(\ref{eq:CMparametric2}). Furthermore, since the modes relevant to $\hat{b_{in}}$ are also supposed to be vacua $|0 \rangle_{b}$, we have the following expression for $\hat{b_{in}}(\omega')\hat{S}_{c}(\zeta)|0 \rangle_{b}$ to the first order in $\zeta$:
\begin{eqnarray}
\hat{b_{in}}(\omega')\hat{S}_{c}(\zeta)|0 \rangle_{b} &=& \hat{b_{in}}(\omega')\big(1+[\zeta^{*} \hat{P}_{\Phi}-\zeta \hat{P}_{\Phi}^{\dagger}]+\frac{1}{2}[\zeta^{*} \hat{P}_{\Phi}-\zeta \hat{P}_{\Phi}^{\dagger}]^{2}+\cdots \big)|0 \rangle_{b} \nonumber \\
&\approx& -\zeta \, \hat{b_{in}}(\omega') \hat{P}_{\Phi}^{\dagger} |0 \rangle_{b} \nonumber \\
&=& -\zeta \, \hat{b_{in}}(\omega') \int_{-\infty}^{\infty} d\omega'_{1} d\omega'_{2} \Phi(\omega'_{1},\omega'_{2}) \hat{a_{in}}^{\dagger}(\omega'_{1})\hat{b_{in}}^{\dagger}(\omega'_{2}) |0 \rangle_{b} \nonumber \\
&=& -\zeta \, \int_{-\infty}^{\infty} d\omega'_{1} d\omega'_{2} \Phi(\omega'_{1},\omega'_{2}) \hat{a_{in}}^{\dagger}(\omega'_{1}) (\hat{b_{in}}^{\dagger}(\omega'_{2})\hat{b_{in}}(\omega')+\delta(\omega'-\omega'_{2})  |0 \rangle_{b} \nonumber \\
&=& -\zeta \, \int_{-\infty}^{\infty} d\omega'_{1} \Phi(\omega'_{1},\omega') \hat{a_{in}}^{\dagger}(\omega'_{1}) |0 \rangle_{b}. \label{eq:singleM1_sv}
\end{eqnarray}
Then, Eq.~(\ref{eq:singleM1}) leads to
\begin{eqnarray}
\langle \hat{N}_{d_{1}} \rangle &=& \frac{1}{2\pi}\eta_{1}|\mathcal{T}|^{2}|\zeta|^{2} \!\int_{-\infty}^{\infty} \!\!\!dt \int_{-\infty}^{\infty} \!\!\!d\omega d\omega'd\omega_{1} d\omega'_{1} \xi^{*}(\omega)\xi(\omega')  \nonumber \\
&\ & \qquad \qquad \qquad \qquad \qquad \times \Phi^{*}(\omega_{1},\omega)\Phi(\omega'_{1},\omega')\big\langle \hat{a_{in}}(\omega_{1}) \hat{a_{in}}^{\dagger}(\omega'_{1}) \big\rangle e^{i(\omega-\omega')t} \nonumber \\
&=& \eta_{1}|\mathcal{T}|^{2}|\zeta|^{2} \int_{-\infty}^{\infty} \!\!\!d\omega d\omega_{1} d\omega'_{1} |\xi(\omega)|^{2} \Phi^{*}(\omega_{1},\omega)\Phi(\omega'_{1},\omega) \big\langle \hat{a_{in}}(\omega_{1}) \hat{a_{in}}^{\dagger}(\omega'_{1}) \big\rangle. \label{eq:singleM1_2}
\end{eqnarray}
Likewise, the number of the photodetection events for the field $\hat{d_{2}}(\omega)$ with detector~2 can be given by
\begin{equation}
\langle \hat{N}_{d_{2}} \rangle = \eta_{2}|\mathcal{R}|^{2}|\zeta|^{2} \int_{-\infty}^{\infty} \!\!\!d\omega d\omega_{1} d\omega'_{1} |\xi(\omega)|^{2} \Phi^{*}(\omega_{1},\omega)\Phi(\omega'_{1},\omega) \big\langle \hat{a_{in}}(\omega_{1}) \hat{a_{in}}^{\dagger}(\omega'_{1}) \big\rangle. \label{eq:singleM2_2}
\end{equation}
Therefore, the normally ordered photodetection for field $\hat{d_{1}}(\omega)$ and that for $\hat{d_{2}}(\omega)$ again coincide with that of the antinormally ordered photodetection for the input signal field $\hat{a_{in}}(\omega)$, although, this time, the frequency-dependent factor, $\Phi^{\dagger}(\omega_{1},\omega)\Phi(\omega'_{1},\omega)$, is added.

The number of the coincidental photodetection events in detector~1 and detector~2 is given by
\begin{eqnarray}
\langle \hat{N}_{d_{1},d_{2}} \rangle &=& \int_{-\infty}^{\infty} dt dt' \langle\hat{d_{1}}^{\dagger}(t) \hat{d_{1}}(t) \hat{d_{2}}^{\dagger}(t') \hat{d_{2}}(t') \rangle  \nonumber \\
&=& \frac{1}{2 \pi} \int_{-\infty}^{\infty} \!\!\!dtdt'\!\int_{-\infty}^{\infty} \!\!\!d\omega d\omega' d\omega'' d\omega''' \xi^{*}(\omega)\xi(\omega')\xi^{*}(\omega'')\xi(\omega''') \nonumber \\
&\ & \qquad \qquad  \times \langle\hat{d_{1}}^{\dagger}(\omega) \hat{d_{1}}(\omega') \hat{d_{2}}^{\dagger}(\omega'') \hat{d_{2}}(\omega''')\rangle e^{i[\omega-\omega']t+i[\omega''-\omega''']t'}. \label{eq:coincidenceM}
\end{eqnarray}
With the vacuum conditions of the modes, $\hat{v}(\omega)$, $\hat{v_{1}}(\omega)$, and $\hat{v_{2}}(\omega)$, the surviving terms in the expansion of Eq.~(\ref{eq:coincidenceM}) are found to be
\begin{eqnarray}
\langle \hat{N}_{d_{1},d_{2}} \rangle &=& \frac{1}{2 \pi} \eta_{1}\eta_{2} \int_{-\infty}^{\infty} \!\!\!dtdt'\!\int_{-\infty}^{\infty} \!\!\!d\omega d\omega' d\omega'' d\omega''' \xi^{*}(\omega)\xi(\omega')\xi^{*}(\omega'')\xi(\omega''') \nonumber \\
&\ & \times  e^{i[\omega-\omega']t+i[\omega''-\omega''']t'}\big( |\mathcal{T}|^{2}|\mathcal{R}|^{2} \langle \hat{b_{out}}^{\dagger}(\omega) \hat{b_{out}}(\omega') \hat{b_{out}}^{\dagger}(\omega'') \hat{b_{out}}(\omega''')\rangle \nonumber \\
&\ & \qquad \qquad \qquad \qquad \qquad \qquad + \mathcal{T}^{*}\mathcal{R}\mathcal{T}^{*}\mathcal{R}\langle \hat{b_{out}}^{\dagger}(\omega) \hat{v}(\omega') \hat{v}^{\dagger}(\omega'') \hat{b_{out}}(\omega''')\rangle \big) \nonumber \\
&=& \frac{1}{2 \pi} \eta_{1}\eta_{2} \int_{-\infty}^{\infty} \!\!\!dtdt'\!\int_{-\infty}^{\infty} \!\!\!d\omega d\omega' d\omega'' d\omega''' \xi^{*}(\omega)\xi(\omega')\xi^{*}(\omega'')\xi(\omega''') \nonumber \\
&\!& \times e^{i[\omega-\omega']t+i[\omega''-\omega''']t'}|\mathcal{T}|^{2}|\mathcal{R}|^{2} \big( \langle \hat{b_{out}}^{\dagger}(\omega) \hat{b_{out}}(\omega') \hat{b_{out}}^{\dagger}(\omega'') \hat{b_{out}}(\omega''')\rangle \nonumber \\
&\ & \qquad \qquad \qquad \qquad \qquad \qquad - \langle \hat{b_{out}}^{\dagger}(\omega) \hat{b_{out}}(\omega''')\rangle \delta(\omega'-\omega'') \big) \nonumber \\
&=& \frac{1}{2 \pi} \eta_{1}\eta_{2}|\mathcal{T}|^{2}|\mathcal{R}|^{2} \int_{-\infty}^{\infty} \!\!\!dtdt'\!\int_{-\infty}^{\infty} \!\!\!d\omega d\omega' d\omega'' d\omega''' \xi^{*}(\omega)\xi(\omega')\xi^{*}(\omega'')\xi(\omega''')  \nonumber \\
&\ & \qquad \qquad \qquad \times \langle \hat{b_{out}}^{\dagger}(\omega) \hat{b_{out}}^{\dagger}(\omega'') \hat{b_{out}}(\omega') \hat{b_{out}}(\omega''')\rangle e^{i[\omega-\omega']t+i[\omega''-\omega''']t'} \nonumber \\
&=& \frac{1}{2 \pi} \eta_{1}\eta_{2}|\mathcal{T}|^{2}|\mathcal{R}|^{2} \int_{-\infty}^{\infty} \!\!\!dtdt'\!\int_{-\infty}^{\infty} \!\!\!d\omega d\omega' d\omega'' d\omega''' \xi^{*}(\omega)\xi(\omega')\xi^{*}(\omega'')\xi(\omega''')  \nonumber \\
&\ & \times \langle \hat{S}_{c}^{\dagger}(\zeta) \hat{b_{in}}^{\dagger}(\omega) \hat{b_{in}}^{\dagger}(\omega'') \hat{b_{in}}(\omega') \hat{b_{in}}(\omega''') \hat{S}_{c}(\zeta)\rangle e^{i[\omega-\omega']t+i[\omega''-\omega''']t'}, \label{eq:coincidenceM_2}
\end{eqnarray}
where we use the relation, $\mathcal{T}\mathcal{R}^{*}=-\mathcal{T}^{*}\mathcal{R}$. Since the modes relevant to $\hat{b_{in}}$ are also supposed to be vacua $|0 \rangle_{b}$, we have the following expression for $\hat{b_{in}}(\omega')\hat{b_{in}}(\omega''')\hat{S}_{c}(\zeta)|0 \rangle_{b}$ to the second order in $\zeta$:
\begin{eqnarray}
&\!& \!\!\! \hat{b_{in}}(\omega')\hat{b_{in}}(\omega''') \hat{S}_{c}(\zeta)|0 \rangle_{b} \nonumber \\
&=& \hat{b_{in}}(\omega')\hat{b_{in}}(\omega''') \big(1+[\zeta^{*} \hat{P}_{\Phi}-\zeta \hat{P}_{\Phi}^{\dagger}]+\frac{1}{2}[\zeta^{*} \hat{P}_{\Phi}-\zeta \hat{P}_{\Phi}^{\dagger}]^{2}+\cdots \big)|0 \rangle_{b} \nonumber \\
&\approx& \frac{1}{2}\zeta^{2} \, \hat{b_{in}}(\omega')\hat{b_{in}}(\omega''') (\hat{P}_{\Phi}^{\dagger})^{2} |0 \rangle_{b} \nonumber \\
&=& \frac{1}{2}\zeta^{2} \, \hat{b_{in}}(\omega')\hat{b_{in}}(\omega''') \int_{-\infty}^{\infty} d\omega'_{1} d\omega'_{2}d\omega'''_{1} d\omega'''_{2} \Phi(\omega'_{1},\omega'_{2})\Phi(\omega'''_{1},\omega'''_{2}) \nonumber \\
&\ & \qquad \qquad \qquad \times \hat{a_{in}}^{\dagger}(\omega'_{1})\hat{b_{in}}^{\dagger}(\omega'_{2})\hat{a_{in}}^{\dagger}(\omega'''_{1})\hat{b_{in}}^{\dagger}(\omega'''_{2}) |0 \rangle_{b} \nonumber \\
&=& \frac{1}{2}\zeta^{2} \, \int_{-\infty}^{\infty} d\omega'_{1} d\omega'_{2}d\omega'''_{1} d\omega'''_{2} \Phi(\omega'_{1},\omega'_{2})\Phi(\omega'''_{1},\omega'''_{2}) \hat{a_{in}}^{\dagger}(\omega'_{1})\hat{a_{in}}^{\dagger}(\omega'''_{1}) \nonumber \\
&\ & \quad \times \big(\delta(\omega'-\omega'_{2})\delta(\omega'''-\omega'''_{2})+\delta(\omega'-\omega'''_{2})\delta(\omega'''-\omega'_{2}) \big) |0 \rangle_{b} \nonumber \\
&=& \frac{1}{2}\zeta^{2} \, \int_{-\infty}^{\infty} d\omega'_{1} d\omega'''_{1} \big( \Phi(\omega'_{1},\omega')\Phi(\omega'''_{1},\omega''') \nonumber \\
&\ & \qquad \qquad \qquad +\Phi(\omega'_{1},\omega''')\Phi(\omega'''_{1},\omega') \big) \hat{a_{in}}^{\dagger}(\omega'_{1})\hat{a_{in}}^{\dagger}(\omega'''_{1})|0 \rangle_{b}, \label{eq:coincidenceM_2_sv}
\end{eqnarray}
where we use the relation:
\begin{eqnarray}
&\ & \big[\,\hat{b_{in}}(\omega')\hat{b_{in}}(\omega'''),\, \hat{b_{in}}^{\dagger}(\omega'_{2})\hat{b_{in}}^{\dagger}(\omega'''_{2})\,\big]|0 \rangle_{b} \nonumber \\
&\ & \quad =\big( \delta(\omega'-\omega'_{2})\delta(\omega'''-\omega'''_{2})+\delta(\omega'-\omega'''_{2})\delta(\omega'''-\omega'_{2}) \big)|0 \rangle_{b}.
\end{eqnarray}
Then, Eq.~(\ref{eq:coincidenceM_2}) leads to
\begin{eqnarray}
\langle \hat{N}_{d_{1},d_{2}} \rangle &=& \frac{1}{(2 \pi)^{2}} \eta_{1}\eta_{2}|\mathcal{T}|^{2}|\mathcal{R}|^{2} \frac{(|\zeta|^{2})^{2}}{4} \int_{-\infty}^{\infty} \!\!\!dtdt'\!\int_{-\infty}^{\infty} \!\!\!d\omega d\omega' d\omega'' d\omega'''\,d\omega_{1} d\omega'_{1} d\omega''_{1} d\omega'''_{1} \nonumber \\
&\ & \qquad \times \xi^{*}(\omega)\xi(\omega')\xi^{*}(\omega'')\xi(\omega''') \nonumber \\
&\ & \qquad \times \big( \Phi^{*}(\omega_{1},\omega)\Phi^{*}(\omega''_{1},\omega'')+\Phi^{*}(\omega_{1},\omega'')\Phi^{*}(\omega''_{1},\omega) \big) \nonumber \\
&\ & \qquad \times \big( \Phi(\omega'_{1},\omega')\Phi(\omega'''_{1},\omega''')+\Phi(\omega'_{1},\omega''')\Phi(\omega'''_{1},\omega') \big) \nonumber \\
&\ & \qquad \times \langle \hat{a_{in}}(\omega_{1})\hat{a_{in}}(\omega''_{1}) \hat{a_{in}}^{\dagger}(\omega'_{1})\hat{a_{in}}^{\dagger}(\omega'''_{1}) \rangle e^{i[\omega-\omega']t+i[\omega''-\omega''']t'} \nonumber \\
&=& \eta_{1}\eta_{2}|\mathcal{T}|^{2}|\mathcal{R}|^{2} \frac{(|\zeta|^{2})^{2}}{4} \int_{-\infty}^{\infty} \!\!\!d\omega d\omega''d\omega_{1} d\omega'_{1} d\omega''_{1} d\omega'''_{1}|\xi(\omega)|^{2}|\xi(\omega'')|^{2} \nonumber \\
&\ & \qquad \times \big( \Phi^{*}(\omega_{1},\omega)\Phi^{*}(\omega''_{1},\omega'')+\Phi^{*}(\omega_{1},\omega'')\Phi^{*}(\omega''_{1},\omega) \big) \nonumber \\
&\ & \qquad \times \big( \Phi(\omega'_{1},\omega)\Phi(\omega'''_{1},\omega'')+\Phi(\omega'_{1},\omega'')\Phi(\omega'''_{1},\omega) \big) \nonumber \\
&\ & \qquad \times \langle \hat{a_{in}}(\omega_{1})\hat{a_{in}}(\omega''_{1}) \hat{a_{in}}^{\dagger}(\omega'_{1})\hat{a_{in}}^{\dagger}(\omega'''_{1}) \rangle. \label{eq:coincidenceM_3}
\end{eqnarray}
As in the case of Eq.~(\ref{eq:Acorrelation}), the surviving contribution to the normally ordered coincidental photodetections for fields $\hat{d_{1}}(\omega)$ and $\hat{d_{2}}(\omega)$ turns out to be made only by the input signal field $\hat{a_{in}}(\omega)$ in antinormal order, though the complicated frequency-dependent factor is added in this continuous-mode counterpart. An attention to the frequency response of the detection (parametric interaction) needs to be paid when the polychromatic-light detection is to be concerned. It should be emphasized that this sort of situation also occurs in measuring the normally ordered correlations\cite{KM1984a}.

Finally, we apply the above continuous-mode treatment to our experiment\cite{UNSTN2004}, in which the antinormally ordered intensity correlations for coherent states were measured by utilizing stimulated parametric down-conversion. Let the input signal field be a continuous-mode coherent state\cite{Loudon} with spectrum amplitude $\alpha(\omega)$:
\begin{equation}
|\alpha \rangle_{a}=\exp\bigl[-\frac{1}{2} \int d\omega |\alpha(\omega)|^{2} \bigr] \sum_{n=0}^{\infty} \frac{\bigl[\int d\omega \alpha(\omega)\hat{a_{in}}^{\dagger}(\omega)\bigr]^{n}}{n!}|0 \rangle_{a}. \label{eq:coherent}
\end{equation}
By rewriting Eqs.~(\ref{eq:singleM1_2}), (\ref{eq:singleM2_2}), and (\ref{eq:coincidenceM_3}) in the normal order, we can exploit the relation, $\hat{a_{in}}(\omega)|\alpha \rangle_{a}=\alpha(\omega)|\alpha \rangle_{a}$. The resultant expressions are respectively written as
\begin{eqnarray}
\langle \hat{N}_{d_{1}} \rangle &=& \eta_{1}|\mathcal{T}|^{2} |\zeta|^{2} \int_{-\infty}^{\infty} \!\!\!d\omega d\omega_{1} d\omega'_{1}|\xi(\omega)|^{2} \nonumber \\
&\ & \qquad \qquad \times \Phi^{*}(\omega_{1},\omega)\Phi(\omega'_{1},\omega) (\alpha^{*}(\omega'_{1})\alpha(\omega_{1})+\delta(\omega'_{1}-\omega_{1}) \nonumber \\
&=& \eta_{1}|\mathcal{T}|^{2} |\zeta|^{2} \int_{-\infty}^{\infty} \!\!\!d\omega d\omega_{1} |\xi(\omega)|^{2} |\Phi(\omega_{1},\omega)|^{2} (\bar{n}+1) \label{eq:singleC1} \\
\langle \hat{N}_{d_{2}} \rangle &=& \eta_{2}|\mathcal{R}|^{2} |\zeta|^{2} \int_{-\infty}^{\infty} \!\!\!d\omega d\omega_{1} d\omega'_{1}|\xi(\omega)|^{2} \nonumber \\
&\ & \qquad \qquad \times \Phi^{*}(\omega_{1},\omega)\Phi(\omega'_{1},\omega) (\alpha^{*}(\omega'_{1})\alpha(\omega_{1})+\delta(\omega'_{1}-\omega_{1}) \nonumber \\
&=& \eta_{2}|\mathcal{R}|^{2} |\zeta|^{2} \int_{-\infty}^{\infty} \!\!\!d\omega d\omega_{1} |\xi(\omega)|^{2} |\Phi(\omega_{1},\omega)|^{2} (\bar{n}+1), \label{eq:singleC2} 
\end{eqnarray}
and
\begin{eqnarray}
&\ & \langle \hat{N}_{d_{1},d_{2}} \rangle = \eta_{1}\eta_{2}|\mathcal{T}|^{2}|\mathcal{R}|^{2} \frac{(|\zeta|^{2})^{2}}{4} \int_{-\infty}^{\infty} \!\!\!d\omega d\omega''d\omega_{1} d\omega'_{1} d\omega''_{1} d\omega'''_{1}|\xi(\omega)|^{2}|\xi(\omega'')|^{2} \nonumber \\
&\ & \qquad\qquad \times \big( \Phi^{*}(\omega_{1},\omega)\Phi^{*}(\omega''_{1},\omega'')+\Phi^{*}(\omega_{1},\omega'')\Phi^{*}(\omega''_{1},\omega) \big) \nonumber \\
&\ & \qquad\qquad \times \big( \Phi(\omega'_{1},\omega)\Phi(\omega'''_{1},\omega'')+\Phi(\omega'_{1},\omega'')\Phi(\omega'''_{1},\omega) \big) \nonumber \\
&\ & \times \Big(\alpha^{*}(\omega''_{1})\alpha^{*}(\omega'''_{1})\alpha(\omega_{1})\alpha(\omega'_{1}) + \delta(\omega_{1}-\omega'''_{1}) \alpha^{*}(\omega''_{1})\alpha(\omega'_{1}) \nonumber \\
&\ & + \delta(\omega'_{1}-\omega'''_{1}) \alpha^{*}(\omega''_{1})\alpha(\omega_{1}) + \delta(\omega_{1}-\omega''_{1}) \alpha^{*}(\omega'''_{1})\alpha(\omega'_{1}) + \delta(\omega_{1}-\omega''_{1})\delta(\omega'_{1}-\omega'''_{1}) \nonumber \\
&\ & \qquad\qquad + \delta(\omega'_{1}-\omega''_{1}) \alpha^{*}(\omega'''_{1})\alpha(\omega_{1}) + \delta(\omega'_{1}-\omega''_{1})\delta(\omega_{1}-\omega'''_{1})\Big) \nonumber \\
&=& \eta_{1}\eta_{2}|\mathcal{T}|^{2}|\mathcal{R}|^{2} (|\zeta|^{2})^{2} \int_{-\infty}^{\infty} \!\!\!d\omega d\omega''d\omega_{1} d\omega''_{1} |\xi(\omega)|^{2}|\xi(\omega'')|^{2} \nonumber \\
&\ & \qquad\qquad \times \Big(|\Phi(\omega_{1},\omega)|^{2}|\Phi(\omega''_{1},\omega'')|^{2} \big(\bar{n}^{2}+2\bar{n}+1) \big) \nonumber \\
&\ & \qquad\qquad +  \Phi^{*}(\omega_{1},\omega)\Phi^{*}(\omega''_{1},\omega'')\Phi(\omega_{1},\omega'')\Phi(\omega''_{1},\omega) \big(2\bar{n}+1) \big) \Big). \label{eq:coincidenceC}
\end{eqnarray}
Here, for simplicity, we approximate $\alpha^{*}(\omega_{1})\alpha(\omega'_{1})$ by $\bar{n}\delta(\omega_{1}-\omega'_{1})$, where $\bar{n}$ describes the average photon number of the input signal field, and likewise the other pairs such as $\alpha^{*}(\omega'_{1})\alpha(\omega''_{1})$ and $\alpha^{*}(\omega''_{1})\alpha(\omega'''_{1})$. These approximations correspond to regarding the input signal field $\hat{a_{in}}(\omega_{1})$ as a stationary and frequency-independent field and may be justified by the fact that the coherence time of the input signal is much longer than that of the down-converted fields and the frequency-dependent characteristics of the input field can be pushed into that of the pump field, i.e., $F_{p}(\omega_{1}+\omega_{2})$. Then, we have the modified antinormally ordered intensity correlation for the continuous-mode field, $\hat{a_{in}}$:
\begin{equation}
\frac{\langle \hat{N}_{d_{1},d_{2}} \rangle}{\langle \hat{N}_{d_{1}} \rangle \langle \hat{N}_{d_{2}} \rangle} = 1+\gamma \big[\, \frac{1}{\bar{n} +1}+\frac{\bar{n}}{[\,\bar{n} +1\,]^{2}} \,\big],  \label{eq:ANHBT_C}
\end{equation}
where
\begin{equation} \label{eq:gamma}
\gamma=  \frac{\int_{-\infty}^{\infty} \!\!\!d\omega d\omega''d\omega_{1} d\omega''_{1} |\xi(\omega)|^{2}|\xi(\omega'')|^{2}\Phi^{*}(\omega_{1},\omega)\Phi^{*}(\omega''_{1},\omega'')\Phi(\omega_{1},\omega'')\Phi(\omega''_{1},\omega)}{\big(\int_{-\infty}^{\infty} \!\!\!d\omega d\omega_{1} |\xi(\omega)|^{2} |\Phi(\omega_{1},\omega)|^{2}\big)^{2}}.
\end{equation}
Here, $1+\gamma$ is exactly equivalent to the form of intensity correlation $g^{(2)}$, for the spontaneous parametric down-conversion, which was derived by Ou \textit{et al.}\cite{ORW1999a}. Physically, value $\gamma$ can be viewed as a measure of the indistinguishability of two emitted photons responsible for a coincidental photodetection. By using a narrow spectral filter, we can eliminate the frequency correlation of the down-converted fields, which gives rise to the distinguishability of two down-converted photon pairs and the reduction of value $\gamma$. Then, since the functions, $\Phi(\omega_{1},\omega)$ etc., can be factorized, the correlation, Eq.~(\ref{eq:ANHBT_C}), reduces to Eq.~(\ref{eq:ANHBT_e}). 

Now, let us move on to the analysis more specific to our experiment. The experimental results\cite{UNSTN2004} were indeed in good agreement with the form, Eq.~({\ref{eq:ANHBT_C}}) with $\gamma=0.45$. To check whether this value of $\gamma$ is really reasonable, we will follow the recent analysis due to de~Riedmatten \textit{et al.}\cite{RSMATZG2003qph}, in which the multiple spontaneous parametric down-conversion process was treated. First, for ease of calculation, we approximate the filter's spectral profile, $|\xi(\omega)|^{2}$, by
\begin{equation} \label{eq:filter}
|\xi(\omega)|^{2}\approx \frac{1}{\sqrt{2\pi\Delta_{F}^{2}}}\exp[-\frac{(\omega-\Omega)^{2}}{2\Delta_{F}^{2}}]
\end{equation}
and the function, $\Phi(\omega_{1},\omega_{2})$, by
\begin{equation} \label{eq:pump}
\Phi(\omega_{1},\omega_{2})\approx \sqrt{F_{p}(\omega_{1}+\omega_{2})} \approx \frac{1}{(2\pi\Delta_{p}^{2})^{\frac{1}{4}}}\exp[-\frac{(\omega_{1}+\omega_{2}-\Omega_{p})^{2}}{4\Delta_{p}^{2}}],
\end{equation}
where $\Delta_{F}^{2}$ and $\Delta_{p}^{2}$ are the variances of the power spectral density for the filter and the pump field, respectively, and $\Omega$ and $\Omega_{p}$ correspond to the center frequency of the filter and pump, respectively. Here, phase-mismatching function, $\phi(\omega_{1},\omega_{2})$, is assumed to be one since the frequency spread of the down-converted fields might be essentially determined by the filter's spectral profile, $|\xi(\omega)|^{2}$, in our experiment. Since both functions, $|\xi(\omega)|^{2}$ and $|\Phi(\omega_{1},\omega_{2})|^{2}$, were assumed to be the normalized Gaussians in Eqs.~(\ref{eq:filter}) and (\ref{eq:pump}), the denominator of Eq.~(\ref{eq:gamma}) becomes one after the integration. The calculation of the numerator in Eq.~(\ref{eq:gamma}) is, on the other hand, rather tedious. Using square completion technique, we can integrate $\phi^{*}(\omega_{1},\omega)\phi(\omega_{1},\omega'')$ with respect to $\omega_{1}$ and $\phi^{*}(\omega''_{1},\omega'')\phi(\omega''_{1},\omega)$ with respect to $\omega''_{1}$. Then, we have the following reduced form of $\gamma$:
\begin{equation} \label{eq:gamma2}
\gamma= \frac{1}{2\pi\Delta_{F}^{2}}\int_{-\infty}^{\infty} \!\!\!d\omega d\omega''\exp[-\frac{\omega^{2}}{2\Delta_{F}^{2}}] \exp[\frac{\omega^{''2}}{2\Delta_{F}^{2}}] \exp[-\frac{1}{4\Delta_{p}^{2}}(\omega^{2}+\omega^{''2}-2\omega\omega'')],
\end{equation}
where, without loss of correctness, we set $\Omega=0$. Using square completion technique again, we first integrate Eq.~(\ref{eq:gamma2}) with respect to $\omega''$ and then with respect to $\omega$. Then, we get
\begin{equation} \label{eq:gamma3_0}
\gamma= \frac{1}{2\pi\Delta_{F}^{2}}\sqrt{2\pi\Gamma^{2}}\sqrt{2\pi\Gamma^{'2}},
\end{equation}
where $\frac{1}{\Gamma^{2}}= \frac{1}{\Delta_{F}^{2}}+ \frac{1}{2\Delta_{p}^{2}}$ and $\frac{1}{\Gamma^{'2}}= \frac{1}{\Gamma^{2}}+ \frac{\Gamma^{2}}{4\Delta_{p}^{4}}$. Thus, the final simplified form of $\gamma$ can be written as\cite{RSMATZG2003qph}
\begin{equation} \label{eq:gamma3}
\gamma= \frac{1}{\sqrt{1+\frac{\Delta_{F}^{2}}{\Delta_{p}^{2}}}}.
\end{equation}
In our experiment, we used the 5-nm-FWHM filter and the 10-nm-FWHM pulse for the pump field. Thus, we have $\gamma \approx 0.9$ from Eq.~(\ref{eq:gamma3}). Besides, since the the filter's spectral profile was found to be not Gaussian, rather Lorentzian, thus the coherence time of the down-converted fields stretched by the filter might be less than that achieved by a filter with the Gaussian spectral profile as was assumed to be in Eq.~(\ref{eq:gamma3}). In addition, the pump pulse duration might be broadened in the course of propagation. Taking account of these factors, value $\gamma$ further reduces by about 10\%. The value is still higher than the observed value of $0.45$. However, Eq.~(\ref{eq:gamma3}) was derived by concerning only the temporal mode mismatching, i.e., the frequency dependence of the down-converted fields. The imperfect alignment, i.e., the spatial mode missmatching, may give rise to the further reduction of the value, $\gamma$. In fact, the coupling efficiency between the down-converted field and the single-mode fiber was 60\% on average. Roughly speaking, this imperfection reduces the value, $\gamma$, by 60\%\cite{ORW1999a}. Thus, the experimental value, $\gamma=0.45$, may be the reasonable end result.

\section{Conclusion and outlook}

We have analyzed a scheme of antinormally ordered photodetector, i.e, quantum counter, which utilized a stimulated parametric down-conversion process, with time-dependent and continuous-mode operators. It has shown that the operator ordering of the intensity correlations measured by such detectors is antinormal even when the measured field has the finite spectrum, though the correlation in general include frequency-dependent factors. The experimentally measured antinormally ordered correlations can be well explained by the present analysis. 

The emission-based antinormally ordered photodetection may provide an interesting alternative for monitoring quantum systems owing to its sensitivity to zero-point fluctuations. The noise reduction in the optical signal amplification\cite{private}, and the realization of the logically as well as \textit{physically}\cite{UIN1996a,Royer1994,KU1999} reversible measurements may be just a few examples of the possible applications of the antinormally ordered photodetection. 


\section*{Acknowledgements}

We thank M.~Hayashi, Y.~Tsuda, T.~Hiroshima, A.~V.~Gopal, M.~Ueda, M.~Kozuma, and H.~Hirayama for their helpful comments.


\end{document}